\documentclass[conference]{IEEEtran}
\usepackage{cite}
\usepackage{amsmath,amssymb,amsfonts}
\usepackage{algorithmic}
\usepackage{graphicx}
\usepackage{textcomp}
\usepackage{xcolor}
\usepackage{multirow}
\usepackage{diagbox}
\usepackage{tikz}
\usepackage{enumitem}
\usepackage{url}
\usepackage{fancyhdr}
\usepackage[bookmarks=true]{hyperref}
\usepackage{mathtools}
\usepackage{color, colortbl}

\hypersetup{
    colorlinks=true,
    linkcolor=blue,
    filecolor=blue,      
    urlcolor=blue,
    citecolor=blue
}

\newcommand\mytab[1]{\begin{tabular}[c]{c}#1\end{tabular}}

\newcommand\rotbox[1]{\rotatebox[origin=c]{90}{\mytab{#1}}}

\newcommand{\empcirc}[2][black,fill=white]{\tikz[baseline=-0.5ex]\draw[#1,radius=#2] (0,0) circle ;} 

\newcommand{\fillcirc}[2][black,fill=black]{\tikz[baseline=-0.5ex]\draw[#1,radius=#2] (0,0) circle ;} 

\def\BibTeX{{\rm B\kern-.05em{\sc i\kern-.025em b}\kern-.08em
    T\kern-.1667em\lower.7ex\hbox{E}\kern-.125emX}}
    
\begin{document}

\title{Modelling Attacks in Blockchain Systems using Petri Nets}

\author{\IEEEauthorblockN{Md. Atik Shahriar\IEEEauthorrefmark{1}, Faisal Haque Bappy\IEEEauthorrefmark{1}, A. K. M. Fakhrul Hossain\IEEEauthorrefmark{1}, Dayamoy Datta Saikat\IEEEauthorrefmark{1},\\Md Sadek Ferdous\IEEEauthorrefmark{1} \IEEEauthorrefmark{2}, Mohammad Jabed M. Chowdhury\IEEEauthorrefmark{3} and  Md Zakirul Alam Bhuiyan \IEEEauthorrefmark{4}}
\IEEEauthorblockA{\IEEEauthorrefmark{1}Department of Computer Science \& Engineering,
Shahjalal University of Science \& Technology, Sylhet, Bangladesh\\}
\IEEEauthorblockA{\IEEEauthorrefmark{2}Imperial College Business School, Imperial College London, London, UK\\}
\IEEEauthorblockA{\IEEEauthorrefmark{3}Department of Computer Science \& Information Technology, La Trobe University, Melbourne, Australia\\}
\IEEEauthorblockA{\IEEEauthorrefmark{4}Department of Computer \& Information Sciences, Fordham University, NY, USA\\
Email: \{md.atikshahriar728, hbfaisal66, a.k.m.fakhrul.hossain, dayamoydatta96\}@gmail.com,\\ sadek-cse@sust.edu, m.chowdhury@latrobe.edu.au, mbhuiyan3@fordham.edu}}

\maketitle

\begin{abstract}
Blockchain technology has evolved through many changes and modifications, such as smart-contracts since its inception in 2008. The popularity of a blockchain system is due to the fact that it offers a significant security advantage over other traditional systems. However, there have been many attacks in various blockchain systems, exploiting different vulnerabilities and bugs, which caused a significant financial loss. Therefore, it is essential to understand how these attacks in blockchain occur, which vulnerabilities they exploit, and what threats they expose. Another concerning issue in this domain is the recent advancement in the quantum computing field, which imposes a significant threat to the security aspects of many existing secure systems, including blockchain, as they would invalidate many widely-used cryptographic algorithms. Thus, it is important to examine how quantum computing will affect these or other new attacks in the future. In this paper, we explore different vulnerabilities in current blockchain systems and analyse the threats that various theoretical and practical attacks in the blockchain expose. We then model those attacks using Petri nets concerning current systems and future quantum computers.
\end{abstract}


\thispagestyle{fancy}
\lhead{This work has been accepted at the 19th IEEE International Conference on Trust, Security and Privacy in Computing and Communications (IEEE TrustCom 2020)}
\cfoot{}

\section{Introduction}

Blockchain technology (or blockchain in short) has emerged as a fundamental technology that offers security through cryptography and consensus mechanisms and addresses single point-of-failure and single point-of-trust issues. The transparency and immutability of the blockchain enable storing publicly provable and indisputable records \cite{rahman2020accountable}. Moreover, the introduction of smart-contracts in blockchain has expanded its utility horizon beyond crypto-currency \cite{al2019privacy}. Indeed, blockchain is being applied in many application domains: from cryptocurrencies to Internet of Things (IoT), healthcare, \& financial systems, supply chain management, and so on.

Although blockchain is considered more secure than many traditional systems, recent reports have shown the security risks correlated with this technology. For example, in the first half of January 2019, Coinbase, a digital currency exchange, observed repeated deep reorganisations of the Ethereum Classic blockchain, most of which contained double-spending worth \$1.1 million \cite{nesbitt_ethereum_2019}. As of February 2018, it has been estimated that hackers have stolen nearly \$2 billion worth of crypto-currency in total since the beginning of 2017 \cite{orcutt_unhackable_2019}. The most (in)famous attack in the history of the blockchain may be the DAO attack on Ethereum stealing more than \$60 million \cite{siegel_dao_2016}. Attackers stole \$450 million from MtGox by exploiting transaction mutability in Bitcoin in March 2014, which caused the Bitcoin trading platform to go bankrupt \cite{adelstein_bitcoin_2016}. \$72 million worth of Bitcoins were stolen from Bitfinex, another currency exchange platform, in 2016 \cite{baldwin_bitfinex_2016}. All those attacks were performed using classical computers. On the other hand, the quantum computing paradigm is advancing quite fast.  Soon, we may get quantum computers more powerful than current supercomputers. Advancements in the quantum computing field will pose a significant threat to the security aspects of many existing secure systems, including blockchain, as they would invalidate many widely-used cryptographic algorithms, creating a new avenue for attackers. To ensure a wide-range adoption of blockchain systems, it is crucial to understand how these attacks in blockchain occur, which vulnerabilities they exploit, and what threats they expose. This paper aims to explore this avenue.

\vspace{1mm}
\noindent \textbf{Contributions.} In this paper, we explore various vulnerabilities that current blockchain systems have and the vulnerabilities that are only exploitable by quantum computers. We also model the associated threats in blockchain by STRIDE, a threat modelling framework, to find out which threats they expose. Lastly, we explore a wide range of attacks, which can be performed both with classical computers and quantum computers, and model them using Petri nets.

\vspace{1mm}
\noindent \textbf{Structure.} In Section \ref{sec:background}, we discuss the background on threat and attack modelling, blockchain, and quantum computing. In Section \ref{sec:relatedWork}, we review some related works. In Section \ref{sec:vulnerabilities}, we present various vulnerabilities within blockchain systems. In Section \ref{sec:attackModelling}, we explore a wide-range of blockchain attacks along with their Petri net models and present the associated threat model using STRIDE. This is followed by a conclusion in Section \ref{sec:conclusion}.

\section{Background}
\label{sec:background}
In this section, we present a brief background on vulnerabilities, threats and attacks (Section \ref{subsec:threats}), threat modelling using STRIDE (Section \ref{subsec:stride}), attack modelling using Petri net (Section \ref{subsec:petri}), different aspects of blockchain (Section \ref{subsec:bc}) and quantum computing (Section \ref{subsec:quantum}). 

\subsection{Vulnerabilities, Threats and Attacks}
\label{subsec:threats}
Vulnerabilities are simply weaknesses in a computer system or network to force a system to act in ways it is not intended to. A threat to a computer system or network is a set of circumstances that can cause a loss or harm by interrupting the operation, functioning, integrity, or availability of the network or system. It can be malicious, accidental, or natural. Humans usually cause malicious and accidental threats.  Attacks are specific techniques or actions deliberately used to harm a system or interrupt the regular services of a computer system or network by exploiting various vulnerabilities. Although threats and attacks both can exploit a system's vulnerabilities, attacks are always intentional, whereas threats may not.

\subsection{Threat modelling}
\label{subsec:stride}
Threat modelling is a technique to help identify and prioritise potential threats, attacks, vulnerabilities, or the absence of appropriate safeguards and countermeasures that can affect a system or network. There are many threat modelling methods. Some well-known threat modelling methods are STRIDE, VAST, PASTA, Trike, Attack tree, and Octave \cite{noauthor_stride_2019}. In this paper, we will only focus on the STRIDE methodology as it has been claimed to be more complete than other models \cite{lipner2006security,johnstone2010threat}. STRIDE, proposed by Loren Kohnfelder and Praerit Garg in 1999 while at Microsoft \cite{kohnfelder1999threats}, is an acronym consisting of first letters from Spoofing, Tampering, Repudiation, Information disclosure, Denial of service, and Elevation of privilege threats. STRIDE helps to identify potential threats in a system that must be mitigated to ensure the system's security.

\subsection{Petri net}
\label{subsec:petri}
\begin{figure}[htbp]
    \setlength\abovecaptionskip{-0.1\baselineskip}
    \includegraphics[width=\linewidth]{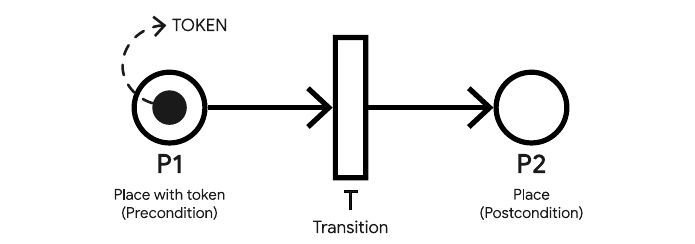}
    \caption{A simple Petri Net}
    \label{fig:petriNet}
\end{figure}

Petri net is a formal mathematical model for studying asynchronous and concurrent processes in distributed systems. It is also known as a place/transition (PT) net. Carl Adam Petri first introduced it in 1962 \cite{petri1966communication}. It is a directed bipartite graph where there are two types of nodes, places, and transitions. Directed arcs connect the places and the transitions and shows which places are preconditions (input) and which places are postconditions (output) after transitions occur. Arcs can only connect places to transitions or transitions to places. Places can hold tokens. Places can store an infinite amount of tokens, but transitions cannot store tokens at all. The state or marking of a Petri net is its distribution of tokens among places. A simple net containing all the elements of a Petri net is shown in Figure \ref{fig:petriNet}, where the circles denote places, and the rectangle denotes a transition.

Formally, a Petri net can be defined as a tuple $N = (P, T, F, M_0)$, where $P$ and $T$ are disjoint finite sets of places and transitions respectively with $P \cap T = \phi$, $F$ is a set of arcs (or incidence function) where $F \subseteq (P \times T) \cup (T \times P)$, and $M_0$ is the initial marking where $M_0 : P \to \{1, 2, 3,...\}$.

The application of Petri nets for attack modelling was first shown by J.P. McDermott as an alternative to attack trees \cite{mcdermott2001attack}. He inspected that Petri nets are better at capturing concurrent operations in the successions of an attack. From then on, Petri nets have been used for modelling both physical and cyber attacks in various systems and networks.

\subsection{Blockchain}
\label{subsec:bc}
Since the early stage of computer technology, computer scientists were trying to create a fully digital and decentralised currency that no central authority can control. Satoshi Nakamoto was the first to solve this classic problem by merging some already existing technologies, such as cryptography, peer-to-peer (P2P) network, and distributed consensus using Proof of Work, to pioneer the idea behind Bitcoin \cite{nakamoto2019bitcoin}, the very first successful digital (crypto-)currency. Satoshi Nakamoto first introduced blockchain as the underlying technology of Bitcoin \cite{nakamoto2019bitcoin}. Blockchain is a distributed, immutable, cryptographically linked, and growing list of a public repository of records where consensus can be established among trustless parties without the interaction of any intermediary. Another revolutionary concept in blockchain is called \textit{Smart-contract}, first proposed by Nick Szabo in the 1990s \cite{szabo1997formalizing} and was introduced in Ethereum by Vitalik Buterin in 2013 \cite{buterin2014next}. Smart-contracts, equipped with a blockchain system, enable immutable, trustless, and transparent distributed computing and autonomous code execution, which has a wide range of applications in different domains. There are two types of blockchains, permissionless or public, where anyone can participate and permissioned or private where only authorised entities can participate. In this research, we will focus on public blockchain systems only. Next, we explore different aspects of a blockchain.

\vspace{1mm}
\noindent \textbf{Transaction and Block: } In a blockchain, a transaction is a signed data structure representing a transfer of a value/data. Transactions are broadcast over a blockchain network and added into the memory pool (called \textit{mempool}). Various special nodes called miners collect and include those transactions into blocks. Blocks are a collection of transactions, tagged with a timestamp, and a hash of the previous block.

\vspace{1mm}
\noindent \textbf{Mining:} Mining is the method in which transactions are verified and added into a block of the blockchain ledger. Mining also produces new crypto-currencies. The special nodes that mine are called \textit{miners}. Mining pools are formed when miners start mining collectively.

\vspace{1mm}
\noindent \textbf{Consensus Mechanism:} Consensus mechanism, a crucial component in any blockchain system, is a mechanism to achieve necessary consensus or agreement on a single value of a data or a particular network state among trustless parties in distributed or multi-agent systems. There are mainly three types of consensus mechanisms used by various public blockchains. They are Proof of Work (PoW), Proof of Stake (PoS), and Delegated Proof of Stake (DPoS) \cite{ferdous2020blockchain}.

\subsection{Quantum Computing}
\label{subsec:quantum}
Unlike classical computers, quantum computers are made of particles (such as superconducting ions, trapped ions) acting as qubits. Because of having a completely different architecture, these particles can use some quantum mechanical phenomena such as superposition and entanglement. A quantum system containing $N$ qubits can generate up to $2^N$ quantum states using superposition and perform operations on all of these $2^N$ states \cite{aaronsonlimits} at a time. However, the classical counterpart can only operate on a single state at a single moment. Hence, the quantum machine can be up to exponentially faster than classical ones in some ways. A crucial quantum algorithm is \textit{Shor's algorithm} \cite{shor1999polynomial}. It can factorise any number exponentially faster than the best known classical algorithm and also can solve discrete logarithm problems. Another popular quantum algorithm is Grover's search algorithm, which can search any intended item from $N$ unsorted items with up to only $\sqrt{N}$ queries and give a quadratic speedup \cite{borbely2007grover}.

\section{Related Work}
\label{sec:relatedWork}

Although there have been some recent studies on blockchain security, very few of them systematically examined the vulnerabilities and threats to blockchain systems and the corresponding real attacks. Saad \textit{et al.} surveyed for papers on blockchain applications and their security vulnerabilities after consulting over 900 research papers in blockchain systems \cite{saad2019exploring}. Li \textit{et al.} conducted a systematic exploration of the security risks and real attacks to popular blockchain systems between 2009 and May 2017 and analysed their corresponding vulnerabilities \cite{li2020survey}. Lee \textit{et al.} analysed the security and vulnerabilities of blockchain systems through a systematic approach \cite{lee2019systematic}. Kabashkin developed a model of risk influence on the effectiveness of blockchain operation \cite{kabashkin2017risk}.

Howard Poston mapped blockchain security threats to STRIDE \cite{poston_threat_nodate}. Sadek \textit{et al.} also argued the resilience of an architecture of a blockchain-based system against frequent security attacks using STRIDE \cite{ferdous2020immutable}. Anna \textit{et al.} described the main attack vectors of blockchain technology in \cite{anna_blockchain_2018}.

Chen \textit{et al.} described how Petri nets can be used to model coordinated cyber-physical attacks \cite{chen2011petri}. Pinna \textit{et al.} introduced a novel approach, based on a Petri net model to analyse the blockchain \cite{pinna2018petri}. However, there is no research that has modelled blockchain attacks using Petri nets.

\section{Vulnerabilities and Attacks}
\label{sec:vulnerabilities}

Various changes, fixes, and reformations have been done over the core concept of blockchain given by Satoshi Nakamoto \cite{nakamoto2019bitcoin}. The more the field expanded, the more vulnerabilities were introduced. Some vulnerabilities are specific to particular blockchain systems, while others are common to every blockchain system. There are mainly six categories of vulnerabilities found in various blockchain systems. 

\subsection{Cryptographic algorithm vulnerabilities}
Most blockchain systems use Elliptic Curve Digital Signature Algorithm (ECDSA) for generating the public-private key pair that ensures the ownership of digital assets. A vulnerability in ECDSA was discovered through which an attacker can recover a user's private key \cite{mayer2016ecdsa}. All elliptic curve cryptography algorithms are based on the assumption that the discrete logarithm problem on an elliptic curve is difficult to solve. Unfortunately, this assumption only holds with any classical computer. However, an attacker with a mature quantum computer may quickly solve this discrete logarithm problem in a short time using Shor's algorithm \cite{shor1999polynomial}. Breaking ECDSA can be used to perform an \textit{Impersonation attack}.

\textit{SHA256} is another cryptographic (hashing) algorithm widely used in most blockchain systems for generating the hash of transactions and blocks. Other popular hash algorithms are Ethash, SCrypt, X11, Equihash, RIPEMD160 \cite{ferdous2020blockchain}. Although these hash functions are secure for classical computers, quantum computers will be able to break many security features of those algorithms. For example, Grover's search algorithm can find collisions of hashes in a feasible time to take down the security of a blockchain \cite{kiktenko2018quantum}. An attacker with a quantum computer will be able to hash faster than any other classical computers and generate blocks much faster than the whole blockchain network. It can lead to \textit{51\% attack}, \textit{Double Spending attack}, or \textit{Selfish-Mining attack}.

\subsection{Consensus mechanism vulnerabilities}
Various consensus mechanisms are used in various blockchain systems. Depending on the consensus mechanism, the verification of transactions and the selection of blocks can be different. The longest chain rule can be exploited for various attacks, such as \textit{51\% attack}, \textit{Selfish Mining attack}, \textit{Double Spending attack}, and \textit{Finney attack} and the GHOST protocol can be exploited for the \textit{Balance attack} \cite{natoli2016balance}.

\subsection{Mining pool vulnerabilities}
Mining pools can bring various degrees of centralisation in a decentralised system like blockchain. Large mining pools control a significant amount of computational power. If large mining pool collude and start mining combinedly, they can perform a \textit{51\% attack} and \textit{Double Spending attack}. Also, mining pools can be a big target for attackers. A miner inside a mining pool can perform a \textit{Block Withholding attack}.

\subsection{Smart-contract vulnerabilities}
Smart-contracts are one of the most vulnerable parts of the blockchain. Nicola \textit{et al.} systematically investigated twelve types of vulnerabilities in smart-contracts \cite{atzei2017survey}. Loi \textit{et al.} found four kinds of potential security bugs in smart-contracts by a symbolic execution tool called Oyente \cite{luu2016making}. Those bugs are timestamp dependence, transaction ordering dependence, reentrancy vulnerability, and mishandled exceptions. The recursive call bug, a reentrancy vulnerability, was exploited to perform the famous \textit{DAO attack} \cite{buterin_dao_2016}, causing Ethereum to be forked and leading to another attack called the \textit{Replay attack}.

\subsection{Design/Architectural vulnerabilities}
Various design faults and architectural vulnerabilities can help attackers to perform an attack. For example, \textit{Eclipse attacks} can take advantage of various design flaws in Ethereum, such as peer's identity, peer selection strategy, inbound vs. outbound connections, and reboot and erase \cite{iotex_eclipse_2018}. \textit{DDOS attack} can also be performed by exploiting the block size limit and mempool flooding.

\subsection{Network vulnerabilities}
Being a P2P network blockchain is highly vulnerable to \textit{Sybil attack}, \textit{Eclipse attack}, and \textit{Block Discarding attack}. Many attacks, such as \textit{Balance attack}, and \textit{Double Spending attack}, take advantage of propagation delays in the network. Many blockchain systems use DNS (Domain Name System) protocols for node discovering, which can be exploited to perform \textit{DNS attack}. Internet traffic can be maliciously diverted to attackers by falsely claiming ownership of IP prefixes to perform the \textit{BGP Hijacking attack}.

\subsection{Summary}
Above, we have identified many attacks in blockchain systems that exploit six vulnerabilities discussed above. We summarise our findings in Table \ref{tab:attacks&Vulnerabilities} where the interrelation between these vulnerabilities and attacks are illustrated. In the table, the symbols ``\fillcirc{3pt}'' and ``\empcirc{3pt}'' have been used to signify if a specific threat is related to a vulnerability or not respectively.

\definecolor{light-gray}{gray}{0.95}
\definecolor{deep-gray}{gray}{0.65}

\begin{table}[hbtp]
\centering
\caption{Blockchain vulnerabilities \& associated attacks}
\label{tab:attacks&Vulnerabilities}
\begin{tabular}{c||c|c|c|c|c|c}
\hline
\rowcolor[gray]{.9} \backslashbox[35mm]{\textbf{Attacks}}{\textbf{Vulnerabilities}} & \rotbox{\textbf{Cryptographic Algorithm}} &
\rotbox{\textbf{Smart-contract}} &
\rotbox{\textbf{Consensus Mechanism}} &
\rotbox{\textbf{Mining Pool}} &
\rotbox{\textbf{Design/Architecture}} &
\rotbox{\textbf{Network}} \\
\hline\hline
\begin{tabular}[c]{@{}c@{}}51\% Attack\end{tabular} & \fillcirc{3pt} & \empcirc{3pt} & \fillcirc{3pt} & \fillcirc{3pt} & \empcirc{3pt} & \empcirc{3pt} \\
\hline
\rowcolor[gray]{.96} \begin{tabular}[c]{@{}c@{}}Impersonation Attack\end{tabular} & \fillcirc{3pt} & \empcirc{3pt} & \empcirc{3pt} & \empcirc{3pt} & \empcirc{3pt} & \empcirc{3pt} \\
\hline
\begin{tabular}[c]{@{}c@{}}Sybil Attack\end{tabular} & \empcirc{3pt} & \empcirc{3pt} & \empcirc{3pt} & \empcirc{3pt} & \empcirc{3pt} & \fillcirc{3pt} \\
\hline
\rowcolor[gray]{.96} \begin{tabular}[c]{@{}c@{}}Eclipse Attack\end{tabular} & \empcirc{3pt} & \empcirc{3pt} & \empcirc{3pt} & \empcirc{3pt} & \fillcirc{3pt} &  \fillcirc{3pt} \\
\hline
\begin{tabular}[c]{@{}c@{}}Selfish-Mining Attack\end{tabular} & \fillcirc{3pt} & \empcirc{3pt} & \fillcirc{3pt} & \fillcirc{3pt} & \empcirc{3pt} & \empcirc{3pt} \\
\hline
\rowcolor[gray]{.96} \begin{tabular}[c]{@{}c@{}}Double Spending Attack\end{tabular} & \fillcirc{3pt} & \empcirc{3pt} & \fillcirc{3pt} & \fillcirc{3pt} & \empcirc{3pt} & \empcirc{3pt} \\
\hline
\begin{tabular}[c]{@{}c@{}}Finney Attack\end{tabular} & \fillcirc{3pt} & \empcirc{3pt} & \fillcirc{3pt}  & \empcirc{3pt} & \empcirc{3pt} & \empcirc{3pt} \\
\hline
\rowcolor[gray]{.96} \begin{tabular}[c]{@{}c@{}}DDOS Attack\end{tabular} & \empcirc{3pt} & \empcirc{3pt} & \empcirc{3pt} & \empcirc{3pt} & \fillcirc{3pt} & \fillcirc{3pt} \\
\hline
\begin{tabular}[c]{@{}c@{}}DNS Attack\end{tabular} & \empcirc{3pt} & \empcirc{3pt} & \empcirc{3pt} & \empcirc{3pt} & \empcirc{3pt} & \fillcirc{3pt} \\
\hline
\rowcolor[gray]{.96} \begin{tabular}[c]{@{}c@{}}BGP Hijacking Attack\end{tabular} & \empcirc{3pt} & \empcirc{3pt} & \empcirc{3pt} & \empcirc{3pt} & \fillcirc{3pt} & \fillcirc{3pt} \\
\hline
\begin{tabular}[c]{@{}c@{}}Block Withholding Attack\end{tabular} & \fillcirc{3pt} & \empcirc{3pt} & \empcirc{3pt} & \fillcirc{3pt} & \empcirc{3pt} & \empcirc{3pt} \\
\hline
\rowcolor[gray]{.96} \begin{tabular}[c]{@{}c@{}}Balance Attack\end{tabular} & \fillcirc{3pt} & \empcirc{3pt} & \fillcirc{3pt} & \fillcirc{3pt} & \empcirc{3pt} & \fillcirc{3pt} \\
\hline
\begin{tabular}[c]{@{}c@{}}Replay Attack\end{tabular} & \empcirc{3pt} & \fillcirc{3pt} & \empcirc{3pt} & \empcirc{3pt} & \fillcirc{3pt} & \empcirc{3pt} \\
\hline
\end{tabular}
\end{table}

\section{Attack \& Threat Modelling}
\label{sec:attackModelling}
In this section we present our attack modelling using Petri nets (Section \ref{subsec:attack}) and threat modelling using STRIDE for the identified attacks (Section \ref{subsec:threatModelling}).
\subsection{Attack Modelling}
\label{subsec:attack}
At first, we model the identified threats using Petri nets. For each attack, we present a brief summary of its method, then present the Petri net and its associated pre and postconditions. For some attacks, we also explain the transitions methods in Petri nets. Transition methods for other attacks are similar and hence excluded for brevity.

\vspace{1mm}
\noindent\textbf{51\% Attack:} PoW based blockchains are particularly susceptible to a 51\% attack  \cite{sayeed2019assessing} which can be launched in two ways. The first and straight forward way is to physically gather more than 50\% computation resources for mining. The concentration of mining power into a few mining pools increases the possibility of this attack. The other way is to use the power of quantum parallelism. Grover's search algorithm using quantum computing makes it possible to search a suitable hash or its collisions using only $\sqrt{N}$ queries in the worst case from $N$ possible hashes, hence, gives a quadratic boost compared to the classical counterparts. This enables an attacker, with a high probability, to mine a block faster than others and consequently to launch a 51\% attack using enough quantum resources \cite{kiktenko2018quantum}. Other PoS and DPoS blockchains are also susceptible to this attack if attackers have enough staking or voting power \cite{sayeed2019assessing}.

\begin{itemize}[leftmargin=1em]
    \item Preconditions:
    \begin{enumerate}[leftmargin=2em,label=$P_\arabic*$]
        \item Attacker has majority hash power
        \begin{enumerate}[label=$P_{1_\alph*}$]
            \item In PoW based blockchains, any of the following:
            \begin{enumerate}[label=$P_{1_{a_\arabic*}}$]
                \item Attacker has more than 50\% hashing power of the entire blockchain.
                \item Attacker has enough quantum resources.
            \end{enumerate}
            \item Attacker has more than 50\% stakes of the total stakes in a PoS based blockchain.
            \item Attacker has more than 50\% voting power in a DPoS based blockchain.
        \end{enumerate}
        \item Attacker has a previous block’s hash.
    \end{enumerate}
    
    \item Transitions:
    \begin{enumerate}[leftmargin=2em,label=$T_\arabic*$]
        \item Generate blocks without broadcasting, thus create an offspring of the blockchain isolated from other blockchain nodes.
        \item Make the isolated offspring of the blockchain longer than the public blockchain by generating blocks more quickly than the whole network.
        \item Broadcast the isolated version of the blockchain to the rest of the network.
    \end{enumerate}
    
    \item Postconditions:
    \begin{enumerate}[leftmargin=2em,label=$P_\arabic*$]
    \setcounter{enumi}{2}
        \item Censor/block transactions.
        \item Hamper usual mining activities of other miners.
        \item Reverse transactions for a Double Spending attack.
        \item Control the market price of cryptocurrencies.
        \item Force other miners to either leave the blockchain or join the mining pool of the attacker.
    \end{enumerate}
    
    \item Petri net for the this attack is presented in Figure \ref{fig:petriNet51Attack}.
    \begin{figure}[htbp]
    \setlength\abovecaptionskip{-0.1\baselineskip}
        \includegraphics[width=\linewidth]{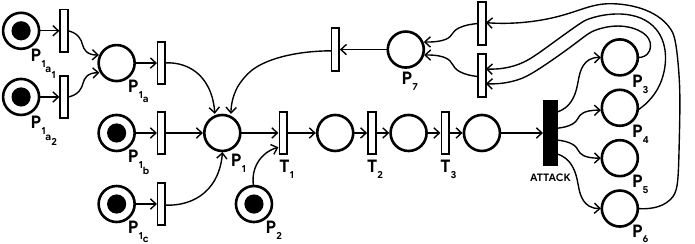}
        \caption{Petri Net of 51\% Attack}
        \label{fig:petriNet51Attack}
    \end{figure}
\end{itemize}

\noindent\textbf{Impersonation Attack:} In this attack, an attacker pretends to be someone else and uses the victim's property to be benefited from it. The attacker mostly targets high officials as victims. Impersonation attacks can be a threat in two ways. The attacker can steal a private key physically or forge a private key by solving the discrete logarithm problem using Shor's Algorithm with quantum resources \cite{roetteler2017quantum} in polynomial time, enabling the attacker to forge a victim's private key of their bitcoin wallet and make transactions.

\begin{itemize}[leftmargin=1em]
    \item Preconditions:
    \begin{enumerate}[leftmargin=2em,label=$P_\arabic*$]
        \item The attacker has enough quantum resource which is capable of solving a discrete logarithm problem.
        \item The attacker has obtained the curve parameters which were used in the ECDSA algorithm to make the public-private key pair.
        \item The attacker may somehow steal the private key physically.
    \end{enumerate}

    \item Postconditions:
    \begin{enumerate}[leftmargin=2em,label=$P_\arabic*$]
    \setcounter{enumi}{3}
        \item Forge the private key of the owner's bitcoin wallet using quantum resources.
        \item Impersonate the owner of the private key by making transactions.
    \end{enumerate}
    
    \item Petri net for this attack is presented in Figure \ref{fig:petriNetImpersonationAttack}.
    \begin{figure}[htbp]
    \setlength\abovecaptionskip{-0.1\baselineskip}
        \includegraphics[width=\linewidth]{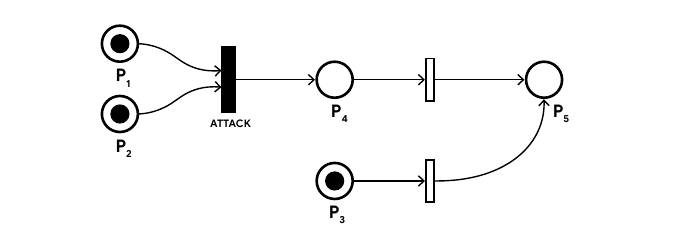}
        \caption{Petri Net of Impersonation Attack}
        \label{fig:petriNetImpersonationAttack}
    \end{figure}
\end{itemize}

\noindent\textbf{Sybil Attack:} The name \textit{Sybil} in Sybil attack was taken from the subject of a 1973 book, Sybil, and was first suggested in 2002 by Brian Zill at Microsoft Research \cite{douceur2002sybil}. In this attack, an attacker creates multiple nodes in a P2P network that appear as different unique nodes to other nodes. In reality, all those nodes belong to a single attacker or a group of attackers. Various blockchains have tried to mitigate the effect of this attack by implementing various consensus mechanisms. Still, a Sybil attack can be possible. Sybil attack also makes way for other attacks, such as Eclipse attack, DDoS attack, and Double Spending attack \cite{zhang2019double}. 

\begin{itemize}[leftmargin=1em]
    \item Preconditions:
    \begin{enumerate}[leftmargin=2em,label=$P_\arabic*$]
        \item The attacker has to create more and more virtual nodes using fake identities.
    \end{enumerate}
    
    \item Postconditions:
    \begin{enumerate}[leftmargin=2em,label=$P_\arabic*$]
    \setcounter{enumi}{1}
        \item Prevent blocks mined by other nodes from propagating into the network by outvoting the honest nodes.
        \item Perform a Double Spending attack by increasing the block propagation time by not sending the new block to other nodes \cite{swathi2019preventing}.
        \item Surround an honest node using the fake nodes and prevent it from connecting to other honest nodes. It can also be done for performing an Eclipse attack.
        \item Conduct a DDoS attack using fake nodes by sending huge amounts of traffic through the network.
    \end{enumerate}
    
    \item Petri net for this attack is presented in Figure \ref{fig:petriNetSybilAttack}.
    \begin{figure}[htbp]
    \setlength\abovecaptionskip{-0.1\baselineskip}
        \includegraphics[width=\linewidth]{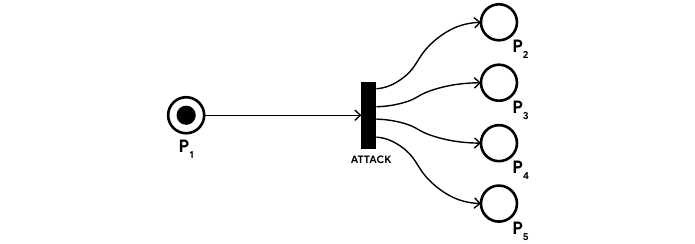}
        \caption{Petri Net of Sybil Attack}
        \label{fig:petriNetSybilAttack}
    \end{figure}
\end{itemize}

\noindent\textbf{Eclipse Attack:} This attack has many similarities with the Sybil attack. However, one significant difference between these attacks is that, while a Sybil attack targets the whole network, Eclipse attack concentrates on a particular node of the network that can either be a miner or just a normal node. This attack obscures and filters a node's view of the entire network by monopolising all of the victim node's incoming and outgoing connections. A lot of other attacks can be executed by taking advantage of this attack, such as 51\% attack, Double Spending attack, and Selfish Mining attack.

\begin{itemize}[leftmargin=1em]
    \item Preconditions:
        \begin{enumerate}[leftmargin=2em,label=$P_\arabic*$]
        \item The attacker has created sufficient amounts of fake nodes. It can be done by a Sybil attack ($P_{1_a}$) or in other ways ($P_{1_b}$).
        \item The attacker has filled in the victim node's internal ``\textit{node address table}'' with invalid addresses and addresses of those fake nodes.
        \item The attacker has made a valid node restart by another attack.
    \end{enumerate}
    
    \item Postconditions:
    \begin{enumerate}[leftmargin=2em,label=$P_\arabic*$]
    \setcounter{enumi}{3}
        \item Filter the view of the whole network for the victim node.
        \item Weaken other competitor miners by eclipsing them.
        \item On a large scale, gain 51\% hashing power by eclipsing other miners.
        \item Perform a 0-confirmation Double Spending attack by only eclipsing the victim node.
        \item Perform an N-confirmation Double Spending attack by eclipsing a miner along with the victim node.
        \item Boost the effort of a selfish miner by deliberately dropping blocks that were found by the eclipsed miners and compete with the blocks found by the attacker.
    \end{enumerate}
    
    \item Petri net for this attack is presented in Figure \ref{fig:petriNetEclipseAttack}.
    \begin{figure}[htbp]
    \setlength\abovecaptionskip{-0.1\baselineskip}
        \includegraphics[width=\linewidth]{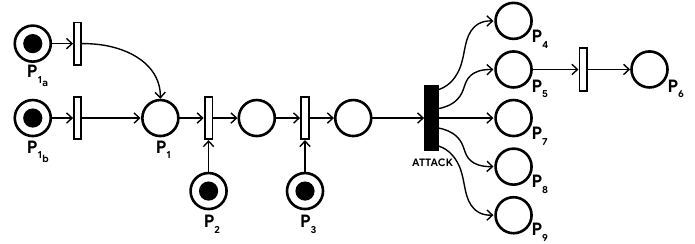}
        \caption{Petri Net of Eclipse Attack}
        \label{fig:petriNetEclipseAttack}
    \end{figure}
\end{itemize}

\noindent\textbf{Selfish-Mining Attack:} This attack is a strategic attack performed by a miner or a mining pool that holds a significant amount of mining power or acquires enough quantum resources to gain a revenue larger than its mining power ratio. A selfish miner or mining pool can invalidate a chain by suddenly introducing a longer chain that the miner kept hidden from the public blockchain network. It has been shown that this attack is one kind of \textit{Block Discarding attack}, and it has been criticised for being not very realistic and not very practical \cite{courtois2014subversive}. A Selfish-Mining attack can facilitate other attacks, such as 51\% attack and Double Spending attack \cite{sapirshtein2016optimal}.

\begin{itemize}[leftmargin=1em]
    \item Preconditions:
    \begin{enumerate}[leftmargin=2em,label=$P_\arabic*$]
        \item The attacker has a previous block's hash.
        \item The attacker has created a private chain forking from the public blockchain by mining blocks privately.
        \item The attacker has made the private chain longer than the public chain ($T_1$) by finding more blocks. It can be possible either by acquiring significant computational power ($P_{3_a}$) or enough quantum resources ($P_{3_b}$).
    \end{enumerate}
    
    \item Transitions:
    \begin{enumerate}[leftmargin=2em,label=$T_\arabic*$]
    \setcounter{enumi}{1}
        \item Not publish the private chain to the network for a strategic amount of time.
        \item Take one of the following paths:
        \begin{enumerate}[label=$T_{3_\alph*}$]
            \item Publish the private chain when the private chain is one single block longer than the public chain.
            \item Withhold the chain for further gain that can result in 3 types of Stubborn-Mining attack ($P_8$) \cite{nayak2016stubborn}.
        \end{enumerate}
    \end{enumerate}
    
    \item Postconditions:
    \begin{enumerate}[leftmargin=2em,label=$P_\arabic*$]
    \setcounter{enumi}{3}
        \item Gain a strategic advantage over other network participants, resulting in revenue more than the attacker's mining power ratio.
        \item Make honest miners wasting their computing power by luring them to keep on working on blocks that leads to a dead-end without gaining any reward. It may force the honest miners to either become selfish themselves or join a selfish mining pool.
        \item At an extreme level, gain 51\% hashing power of the blockchain network.
        \item For every successful Selfish-Mining attack, there is a high probability of successfully launching a Double Spending attack by the selfish miner \cite{sapirshtein2016optimal}.
    \end{enumerate}
    \item Petri net for this attack is presented in Figure \ref{fig:petriNetSelfishMiningeAttack}.
    \begin{figure}[htbp]
    \setlength\abovecaptionskip{-0.1\baselineskip}
        \includegraphics[width=\linewidth]{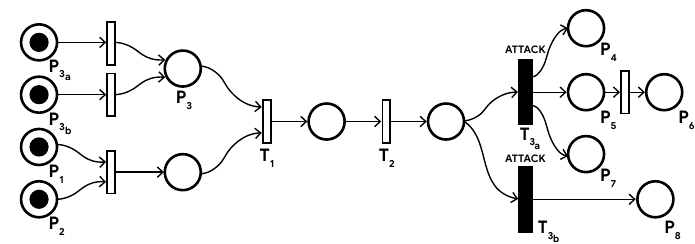}
        \caption{Petri Net of Selfish-Mining Attack}
        \label{fig:petriNetSelfishMiningeAttack}
    \end{figure}
\end{itemize}

\noindent\textbf{Double Spending Attack:} One of the main goals of the consensus mechanism in blockchain is to ensure that a crypto-currency cannot be duplicated digitally and spent twice or more. However, in certain circumstances, it has been proved that spending a currency twice or more can still be possible, referred to as Double Spending (DS) attack. There are two variants of this attack, 0-confirmation DS attack, and N-confirmation DS attack.

The 0-confirmation DS attack, also known as \textit{Race attack}, is likely to happen in fast payment systems where a merchant releases a product to a consumer for a transaction that has not been confirmed yet by the blockchain network \cite{karame2012double}. The attacker performs this attack with the help of one or more helpers, and it has a high probability of succeeding even if the attacker has no computation power at all.

The N-confirmation attack is difficult to perform as it requires the possession of a significant amount of computational power or acquisition of enough quantum resources for modifying N blocks of the blockchain network. This attack depends on the value of N and the attacker's computational power. Nevertheless, even if the attacker has less computational power, there is always a probability of succeeding in the attack \cite{rosenfeld2014analysis}. Other attacks such as 51\% attack, Selfish-Mining attack, Eclipse attack, and Sybil attack can significantly improve the probability of this attack.

\begin{itemize}[leftmargin=1em]
    \item Preconditions:
    \begin{enumerate}[leftmargin=2em,label=$P_\arabic*$]
        \item For 0-confirmation DS attack -
        \begin{enumerate}[label=$P_{1_\alph*}$]
            \item A transaction is received by the victim node.
            \item Another transaction conflicting with the previous transaction is confirmed by the blockchain network with the help of some helpers.
            \item Before the attack is detected, the service of the victim is received by the attacker.
        \end{enumerate}
        \item For N-confirmation DS attack -
        \begin{enumerate}[label=$P_{2_\alph*}$]
            \item The attacker has acquired enough computational power ($P_{2_{a_1}}$) or quantum resources ($P_{2_{a_2}}$).
            \item A transaction is sent to the victim. After the victim got N-confirmation, the attacker has published a private chain that does not include the previous transaction and that is longer than N ($T_1$).
        \end{enumerate}
    \end{enumerate}
    \vspace{3mm}
    \item Postconditions:
    \begin{enumerate}[leftmargin=2em,label=$P_\arabic*$]
    \setcounter{enumi}{2}
        \item Spend the same digital currency twice or more.
    \end{enumerate}
    
    \item Petri net for this attack is presented in Figure \ref{fig:petriNetDoubleSpendingAttack}.
    \begin{figure}[htbp]
    \setlength\abovecaptionskip{-0.1\baselineskip}
        \includegraphics[width=\linewidth]{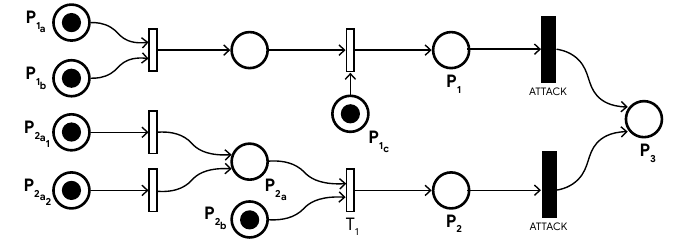}
        \caption{Petri Net of Double Spending Attack}
        \label{fig:petriNetDoubleSpendingAttack}
    \end{figure}
\end{itemize}

\noindent\textbf{Finney Attack:} This attack \cite{hal_finney_2011} is a variant of the 0-confirmation DS attack and is complicated to use in practice as it requires a time-sensitive procedure. The lower the attacker's hash rate, the less chance he has of carrying out the attack. If the attack is intended to obtain any illiquid good, it is hard to manage the need for this good \cite{bitstack_finney_nodate}. Acquisition of enough quantum resources can facilitate the attacker.

\begin{itemize}[leftmargin=1em]
    \item Preconditions:
    \begin{enumerate}[leftmargin=2em,label=$P_\arabic*$]
        \item The attacker has made a transaction from one address to another address, both controlled by the attacker.
        \item The attacker has mined a block that includes the previous transaction without broadcasting the transaction. It can be done either by having a certain computational power ($P_{2_a}$) or acquiring enough quantum resources ($P_{2_b}$). The block is not broadcast if found.
        \item The attacker has made the same transaction again from the attacker's address to a merchant's address.
        \item The merchant accepts the unconfirmed second transaction. Then, the attacker publishes the previously mined block ($T_1$).
    \end{enumerate}
    
    \item Postconditions:
    \begin{enumerate}[leftmargin=2em,label=$P_\arabic*$]
    \setcounter{enumi}{4}
        \item Invalidate an uncorfirmed transaction.
        \item Spend the same crypto-currency twice.
    \end{enumerate}
    
    \item Petri net for this attack is presented in Figure \ref{fig:petriNetFinneyAttack}.
    \begin{figure}[htbp]
    \setlength\abovecaptionskip{-0.1\baselineskip}
        \includegraphics[width=\linewidth]{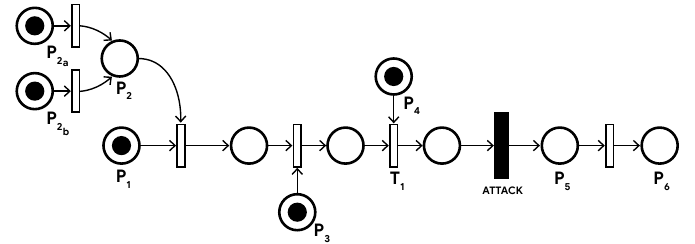}
        \caption{Petri Net of Finney Attack}
        \label{fig:petriNetFinneyAttack}
    \end{figure}
\end{itemize}

\noindent\textbf{DDOS Attack:} Distributed Denial of Service (DDoS) attack is used for a complete or partial service disruption. Two types of DDoS attacks can be found on blockchains. In the classical attack, the attacker tries to exploit the limitation in block size and generate dust (spam) transactions that obtain the space and prevent other transactions from being mined \cite{baqer2016stressing}. The other DDoS attack targets mempools by suffocating them with a flood of unconfirmed transactions.

\begin{itemize}[leftmargin=1em]
    \vspace{2mm}
    \item Preconditions:
    \begin{enumerate}[leftmargin=2em,label=$P_\arabic*$]
        \item The attacker has exploited the block size limit to overwhelm blocks with low-valued spam transactions.
        \item The transactions have a greater-than-zero age (confirmation score) that is enough to pay the relay and mining fee.
        \item The exchange rate of transactions is higher than the network output.
    \end{enumerate}
    
    \item Postconditions:
    \begin{enumerate}[leftmargin=2em,label=$P_\arabic*$]
    \setcounter{enumi}{3}
        \item Cause delay in the verification of legitimate transactions.
        \item Stop crypto-currency circulation or block processing for a while.
        \item Make crypto-currency vulnerable to flood attacks.
        \item Flood mempools with unconfirmed dust transactions.
        
    \end{enumerate}
    
    \item Petri net for this attack is presented in Figure \ref{fig:petriNetDDOSAttack}.
    \begin{figure}[htbp]
    \setlength\abovecaptionskip{-0.1\baselineskip}
        \includegraphics[width=\linewidth]{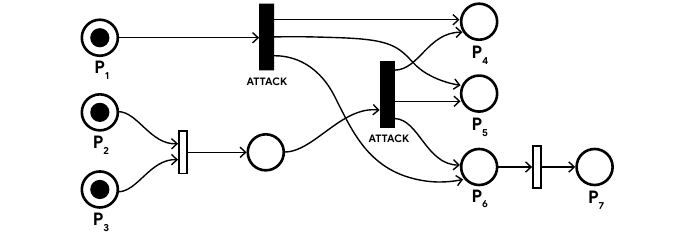}
        \caption{Petri Net of DDOS Attack}
        \label{fig:petriNetDDOSAttack}
    \end{figure}
\end{itemize}

\noindent\textbf{DNS Attack:} Resolution of the Domain Name System (DNS) may trigger vulnerabilities in a blockchain-based system for DDoS attack, Man-in-the-middle attack (at the resolver side), cache poisoning, and old records \cite{saad2019exploring}. When a new node is connected to the network, the active peers (identified by IP addresses) need to be discovered through a bootstrapping mechanism. DNS can be used as a bootstrapping mechanism for querying the active nodes. To launch this attack, an attacker may either insert an invalid list of seed nodes into the open-source blockchain program or poison the resolver's DNS cache.

\begin{itemize}[leftmargin=1em]
    \item Preconditions:
    \begin{enumerate}[leftmargin=2em,label=$P_\arabic*$]
        \item DNS is used as the bootstrapping mechanism.
        \item DNS caches have been poisoned by the attacker.
    \end{enumerate}
    
    \item Postconditions:
    \begin{enumerate}[leftmargin=2em,label=$P_\arabic*$]
    \setcounter{enumi}{2}
        \item Make Bitcoin or other crypto-currencies vulnerable to many different attacks like Man-in-the-middle attack, DDoS attack.
        \item Isolate blockchain peers and divert them to fabricated networks.
    \end{enumerate}
    
    \item Petri net for this attack is presented in Figure \ref{fig:petriNetDNSAttack}.
    \begin{figure}[htbp]
    \setlength\abovecaptionskip{-0.1\baselineskip}
        \includegraphics[width=\linewidth]{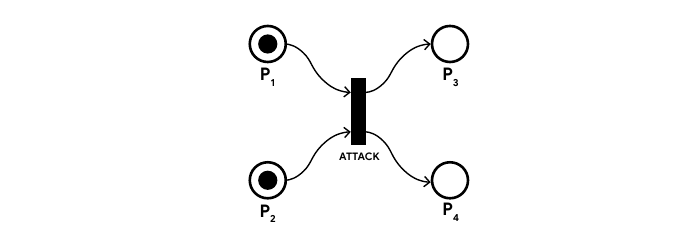}
        \caption{Petri Net of DNS Attack}
        \label{fig:petriNetDNSAttack}
    \end{figure}
\end{itemize}

\noindent\textbf{BGP Hijacking Attack:} BGP (Border Gateway Protocol) hijacking is done by falsely claiming ownership of IP address groups, called IP prefixes, which may not be owned, regulated, or redirected to and allows the attackers to divert Internet traffics maliciously. Internet Service Providers (ISPs) controls the traffic flows on the internet as they own one or more Autonomous Systems (ASes) which handle the traffic routing \cite{gao2001inferring}. Full nodes, nodes that maintain a full copy of the network state, are distributed spatially within an AS or ISP over the Internet  \cite{saad2019exploring}, and they become vulnerable to the BGP hijacking attack. An attacker can hijack the traffic of a target AS where the majority of the blockchain nodes are hosted.

\begin{itemize}[leftmargin=1em]
    \item Preconditions:
    \begin{enumerate}[leftmargin=2em,label=$P_\arabic*$]
        \item The nodes are spatially spread over an AS or ISP.
        \item The attacker has announced a smaller range of IP addresses than other ASes.
        \item The attacker has offered a shorter route to certain blocks of IP addresses. 
    \end{enumerate}
    
    \item Postconditions:
    \begin{enumerate}[leftmargin=2em,label=$P_\arabic*$]
    \setcounter{enumi}{3}
        \item Reduce the hash rate of the blockchain system.
        \item Block propagation can be delayed by up to 20 minutes.
        \item Increase the possibility of other attacks such as Double Spending Attack, Balance, Consensus delay, and Blockchain fork.
    \end{enumerate}
    
    \item Petri net for this attack is presented in Figure \ref{fig:petriNetBGPHijackingAttack}.
    \begin{figure}[htbp]
    \setlength\abovecaptionskip{-0.1\baselineskip}
        \includegraphics[width=\linewidth]{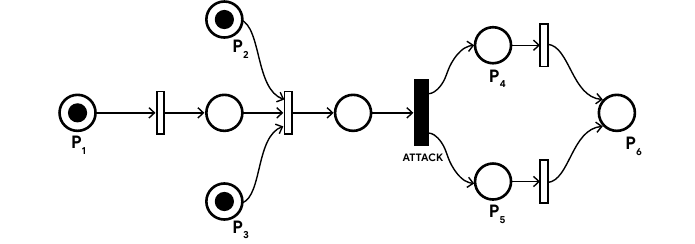}
        \caption{Petri Net of BGP Hijacking Attack}
        \label{fig:petriNetBGPHijackingAttack}
    \end{figure}
\end{itemize}

\noindent\textbf{Block Withholding Attack:} This attack can only be executed by a miner who is a part of a mining pool based on PoW consensus. In this attack, the dishonest miner (attacker), submits all shares to the operator but does not submit valid blocks, if found, to the operator. Instead, the attacker withholds the blocks for two types of attack, Sabotage, and Lie in Wait \cite{rosenfeld2011analysis}. In the Sabotage attack, the attacker does not submit any block at all. However, in the Lie in Wait attack, the attacker postpones submitting the blocks for some time and uses them to mine where the reward is most \cite{rosenfeld2011analysis}. It is claimed in \cite{courtois2014subversive} that the Sabotage attack can be profitable for the attacker. The attacker can use quantum resources to increase the probability of finding blocks.

\begin{itemize}[leftmargin=1em]
    \item Preconditions:
    \begin{enumerate}[leftmargin=2em,label=$P_\arabic*$]
        \item The attacker is a miner of a mining pool. The attacker mines usually and submits all shares to the pool operator for gaining trust.
        \item The attacker has found a block either by normal mining ($P_{2_a}$) or by using enough quantum resources ($P_{2_b}$).
    \end{enumerate}
    
    \item Transitions can be any of the following two:
    \begin{enumerate}[leftmargin=2em,label=$T_\arabic*$]
        \item Either not submit the block at all to perform a Sabotage attack.
        \item Use the block to mine where the reward is most to perform a Lie in Wait attack.
    \end{enumerate}
    
    \item Postconditions:
    \begin{enumerate}[leftmargin=2em,label=$P_\arabic*$]
    \setcounter{enumi}{2}
        \item Earn more from mining than usual.
        \item Harm the mining pool by depriving it of getting the reward of blocks.
        \item On a large scale, destroy a mining pool.
    \end{enumerate}
    
    \item Petri net for this attack is presented in Figure \ref{fig:petriNetBlockWithholdingAttack}.
    \begin{figure}[htbp]
    \setlength\abovecaptionskip{-0.1\baselineskip}
        \includegraphics[width=\linewidth]{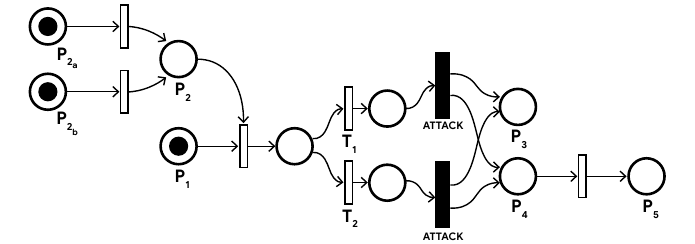}
        \caption{Petri Net of Block Withholding Attack}
        \label{fig:petriNetBlockWithholdingAttack}
    \end{figure}
\end{itemize}

\noindent\textbf{Balance Attack:} Ethereum uses a modified version of Greedy Heaviest-Observed Sub-Tree (GHOST) \cite{buterin2014next}. This creates ground for a new type of attack called Balance attack. In this attack, an attacker delays communications between subgroups of similar mining power for some time to double-spend successfully \cite{natoli2016balance}. It is evident that to execute this attack successfully, the attacker needs both a significant hashing power (or enough quantum resources) and delay time \cite{natoli2016balance}.

\begin{itemize}[leftmargin=1em]
    \item Preconditions:
    \begin{enumerate}[leftmargin=2em,label=$P_\arabic*$]
        \item The attacker has delayed communication between two subgroups of similar mining power.
        \item The attacker has issued a transaction in one subgroup where the service provider belongs. The transaction should get enough confirmation to convince the provider of its validity.
        \item The attacker continues mining in another subgroup and tries to create a chain that outweighs the chain mined by the previous subgroup. He may have significant hashing power ($P_{3_a}$) or quantum resources ($P_{3_b}$).
    \end{enumerate}
    
    \item Postconditions:
    \begin{enumerate}[leftmargin=2em,label=$P_\arabic*$]
    \setcounter{enumi}{3}
        \item Deliberately influence the branch selection process.
        \item Invalidate a confirmed transaction.
        \item Spend the same crypto-currency twice.
    \end{enumerate}
    
    \item Petri net for this attack is presented in Figure \ref{fig:petriNetBalanceAttack}.
    \begin{figure}[htbp]
    \setlength\abovecaptionskip{-0.1\baselineskip}
        \includegraphics[width=\linewidth]{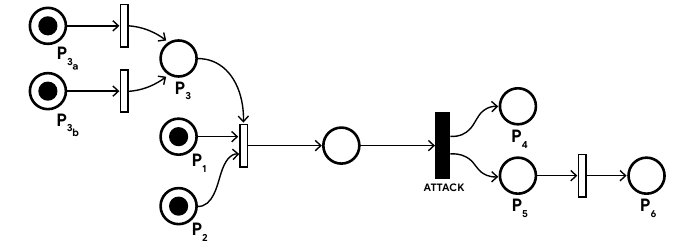}
        \caption{Petri Net of Balance Attack}
        \label{fig:petriNetBalanceAttack}
    \end{figure}
\end{itemize}

\noindent\textbf{Replay Attack:} A replay attack happens when an attacker sniffs a packet from one blockchain and replays it to another blockchain. As a result, the victim loses assets in both chains. If the packet is an authenticated packet, the replay attack will enable the attacker to authenticate as someone else and, subsequently, access another person's resources or data \cite{abdullah2017blockchain}. A replay attack does not mean that anyone else has control over a user's money. It will only clone an existing transaction from the recently forked blockchain and render the same one in the old blockchain.

\begin{itemize}[leftmargin=1em]
    \item Preconditions:
    \begin{enumerate}[leftmargin=2em,label=$P_\arabic*$]
        \item A hard fork has occurred in the chain to generate two chains sharing the exact same transaction history.
        \item The attacker has copied some transactions from the old chain and broadcast in the new chain. 
    \end{enumerate}
    
    \item Postconditions:
    \begin{enumerate}[leftmargin=2em,label=$P_\arabic*$]
    \setcounter{enumi}{2}
        \item One transaction is validated twice in the blockchain.
        \item User loses his assets in both old and new chains.
    \end{enumerate}
    
    \item Petri net for this attack is presented in Figure \ref{fig:petriNetReplayAttack}.
    \begin{figure}[htbp]
    \setlength\abovecaptionskip{-0.1\baselineskip}
        \includegraphics[width=\linewidth]{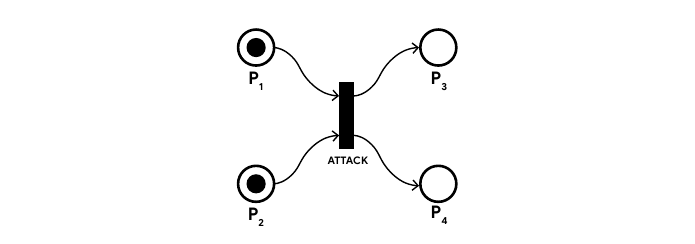}
        \caption{Petri Net of Replay Attack}
        \label{fig:petriNetReplayAttack}
    \end{figure}
\end{itemize}
\begin{table}[hbtp]
\centering
\caption{Correlation of attacks, incentives and quantum effect}
\label{tab:attacks}
\begin{tabular}{l||l|l|c}
\hline
\rowcolor[gray]{.9} \multicolumn{1}{c||}{\rotbox{\textbf{Attacks}}} & \multicolumn{1}{c}{\rotbox{\textbf{Influenced}\\\textbf{Attacks}}} & \multicolumn{1}{|c}{\rotbox{\textbf{Possible}\\\textbf{Motivations}}} & \multicolumn{1}{|c}{\rotbox{\textbf{Quantum}\\\textbf{Effect}}} \\
\hline
\hline
1. 51\% Attack & $\Rightarrow$ 6 & \begin{tabular}[c]{@{}l@{}}Financial gain\\Harm others\\Harm to the system\end{tabular} & \fillcirc{3pt} \\
\hline
\rowcolor[gray]{.96} 2. Impersonation Attack & & \begin{tabular}[c]{@{}l@{}}Financial gain\\Harm others\end{tabular} & \fillcirc{3pt} \\
\hline
3. Sybil Attack & \begin{tabular}[c]{@{}l@{}}$\Rightarrow$ 4\\$\Rightarrow$ 6\\$\Rightarrow$ 8\end{tabular} & \begin{tabular}[c]{@{}l@{}}Financial gain\\Harm others\\Harm to the system\end{tabular} & \empcirc{3pt} \\
\hline
\rowcolor[gray]{.96} 4. Eclipse Attack & \begin{tabular}[c]{@{}l@{}}$\Rightarrow$ 6\\$\Rightarrow$ 5\\$\Rightarrow$ 1\end{tabular} & \begin{tabular}[c]{@{}l@{}}Financial gain\\Harm others\end{tabular} & \empcirc{3pt} \\
\hline
5. Selfish-Mining Attack & \begin{tabular}[c]{@{}l@{}}$\Rightarrow$ 6\\$\Rightarrow$ 1\end{tabular} & \begin{tabular}[c]{@{}l@{}}Financial gain\\Harm to the system\end{tabular} & \fillcirc{3pt} \\
\hline
\rowcolor[gray]{.96} 6. Double Spending Attack & & \begin{tabular}[c]{@{}l@{}}Financial gain\\Harm others\end{tabular} & \fillcirc{3pt} \\
\hline
7. Finney Attack & $\Rightarrow$ 6 & \begin{tabular}[c]{@{}l@{}}Financial gain\\Harm others\end{tabular} & \fillcirc{3pt} \\
\hline
\rowcolor[gray]{.96} 8. DDoS Attack & & Harm to the system & \empcirc{3pt} \\
\hline
9. DNS Attack & $\Rightarrow$ 8 & Harm to the system & \empcirc{3pt} \\
\hline
\rowcolor[gray]{.96} 10. BGP Hijacking Attack & \begin{tabular}[c]{@{}l@{}}$\Rightarrow$ 8\\$\Rightarrow$ 6\\$\Rightarrow$ 12\end{tabular} & \begin{tabular}[c]{@{}l@{}}Financial gain\\Harm others\\Harm to the system\end{tabular} & \empcirc{3pt} \\
\hline
11. Block Withholding Attack & & \begin{tabular}[c]{@{}l@{}}Financial gain\\Harm to the system\end{tabular} & \fillcirc{3pt} \\
\hline
\rowcolor[gray]{.96} 12. Balance Attack & $\Rightarrow$ 6 & \begin{tabular}[c]{@{}l@{}}Financial gain\\Harm others\end{tabular} & \fillcirc{3pt} \\
\hline
13. Replay Attack & $\Rightarrow$ 6 & \begin{tabular}[c]{@{}l@{}}Financial gain\\Harm others\end{tabular} & \empcirc{3pt} \\
\hline
\end{tabular}
\end{table}

\noindent \textbf{Summary:}
As one attack in blockchain systems can lead to another attack, there is a correlation among them. Another vital thing about attacks is the incentives behind them. Attackers attack mainly for their personal gain, but it may not be true all the time. Also, it is clear that not all attacks are impacted by quantum computing. To clarify these issues, we summarise the correlation of various attacks, motivations, and their quantum effects in Table \ref{tab:attacks}, with symbols ``\fillcirc{3pt}'' and ``\empcirc{3pt}'' implying impacted by a quantum effect or not respectively. According to the table, Sybil, Eclipse, and BGP attacks have influenced the highest number of attacks: three. For example, Sybil attack has influenced Eclipse, DDoS and Double Spending attacks, and so on. The principal motivations of these attacks are financial gains; however, attacks are also launched to cause harm to persons or systems. Finally, seven out of thirteen, more than $50\%$ of attacks, are impacted by quantum computing. 

\subsection{Threat Modelling}
\label{subsec:threatModelling}
Every attack in any system exposes some threats, which is true for any blockchain system as well. To model the threats from the identified attacks, we have utilised STRIDE. The modelled threats against the attacks are presented in Table \ref{tab:threatModelling} with symbols ``\fillcirc{3pt}'' and ``\empcirc{3pt}'' implying to have a correlation between an attack and threat and no correlation respectively.

\begin{table}[htbp]
\centering
\caption{STRIDE threat modelling of attacks in blockchain}
\label{tab:threatModelling}
\begin{tabular}{l||c|c|c|c|c|c}
\hline
\rowcolor[gray]{.9} \textbf{Attacks} & \textbf{S} & \textbf{T} & \textbf{R} & \textbf{I} & \textbf{D} & \textbf{E} \\
\hline \hline
\begin{tabular}[c]{@{}c@{}}51\% Attack\end{tabular} & \empcirc{3pt} & \fillcirc{3pt} & \empcirc{3pt} & \empcirc{3pt} & \fillcirc{3pt} & \empcirc{3pt} \\
\hline
\rowcolor[gray]{.96} \begin{tabular}[c]{@{}c@{}}Impersonation Attack\end{tabular} & \fillcirc{3pt} & \fillcirc{3pt} & \fillcirc{3pt} & \fillcirc{3pt} & \empcirc{3pt} & \fillcirc{3pt} \\
\hline
\begin{tabular}[c]{@{}c@{}}Sybil Attack\end{tabular} & \fillcirc{3pt} & \empcirc{3pt} & \empcirc{3pt} & \empcirc{3pt} & \fillcirc{3pt} & \empcirc{3pt} \\
\hline
\rowcolor[gray]{.96} \begin{tabular}[c]{@{}c@{}}Eclipse Attack\end{tabular} & \fillcirc{3pt} & \fillcirc{3pt} & \fillcirc{3pt} & \empcirc{3pt} & \empcirc{3pt} & \fillcirc{3pt} \\
\hline
\begin{tabular}[c]{@{}c@{}}Selfish-Mining Attack\end{tabular} & \empcirc{3pt} & \fillcirc{3pt} & \empcirc{3pt} & \empcirc{3pt} & \empcirc{3pt} & \empcirc{3pt} \\
\hline
\rowcolor[gray]{.96} \begin{tabular}[c]{@{}c@{}}Double Spending  Attack\end{tabular} & \empcirc{3pt} & \fillcirc{3pt} & \empcirc{3pt} & \empcirc{3pt} & \empcirc{3pt} & \empcirc{3pt} \\
\hline
\begin{tabular}[c]{@{}c@{}}Finney Attack\end{tabular} & \empcirc{3pt} & \fillcirc{3pt} & \empcirc{3pt} & \empcirc{3pt} & \empcirc{3pt} & \empcirc{3pt} \\
\hline
\rowcolor[gray]{.96} \begin{tabular}[c]{@{}c@{}}DDOS Attack\end{tabular} & \empcirc{3pt} & \empcirc{3pt} & \empcirc{3pt} & \empcirc{3pt} & \fillcirc{3pt} & \fillcirc{3pt} \\
\hline
\begin{tabular}[c]{@{}c@{}}DNS Attack\end{tabular} & \empcirc{3pt} & \fillcirc{3pt} & \empcirc{3pt} & \empcirc{3pt} & \empcirc{3pt} & \empcirc{3pt} \\
\hline
\rowcolor[gray]{.96} \begin{tabular}[c]{@{}c@{}}BGP Hijacking Attack\end{tabular} & \empcirc{3pt} & \empcirc{3pt} & \empcirc{3pt} & \empcirc{3pt} & \fillcirc{3pt} & \empcirc{3pt} \\
\hline
\begin{tabular}[c]{@{}c@{}}Block Withholding Attack\end{tabular} & \empcirc{3pt} & \empcirc{3pt} & \fillcirc{3pt} & \empcirc{3pt} & \fillcirc{3pt} & \fillcirc{3pt} \\
\hline
\rowcolor[gray]{.96} \begin{tabular}[c]{@{}c@{}}Balance Attack\end{tabular} & \empcirc{3pt} & \fillcirc{3pt} & \empcirc{3pt} & \empcirc{3pt} & \empcirc{3pt} & \empcirc{3pt} \\
\hline
\begin{tabular}[c]{@{}c@{}}Replay Attack\end{tabular} & \fillcirc{3pt} & \fillcirc{3pt} & \empcirc{3pt} & \empcirc{3pt} & \empcirc{3pt} & \empcirc{3pt} \\
\hline
\end{tabular}
\end{table}

As per the table, all identified attacks have a direct impact on at least one STRIDE threat. Among the attacks, Impersonation exposes the highest threats, five out of six, whereas Selfish-Mining, Double Spending, Finney, DNS, and Balance attacks expose the lowest threat: one. On the contrary, tampering is the most exposed threat by those attacks.

\section{Conclusion}
\label{sec:conclusion}

In this paper, we have explored several attacks for blockchain systems using classical and quantum computing. We have modelled those attacks using Petri nets and modelled the associated threats using STRIDE to illustrate their impact relations with those attacks. We have also analysed the main vulnerabilities that typical blockchain systems have. We believe that, blockchain in the quantum computing era will be much more vulnerable. Therefore, more research should be carried towards that direction. In future, we will focus on private blockchain systems which have been excluded in this research.

\bibliographystyle{IEEEtran}
\bibliography{arXivSubmission}

\begin{thebibliography}{10}
\providecommand{\url}[1]{#1}
\csname url@samestyle\endcsname
\providecommand{\newblock}{\relax}
\providecommand{\bibinfo}[2]{#2}
\providecommand{\BIBentrySTDinterwordspacing}{\spaceskip=0pt\relax}
\providecommand{\BIBentryALTinterwordstretchfactor}{4}
\providecommand{\BIBentryALTinterwordspacing}{\spaceskip=\fontdimen2\font plus
\BIBentryALTinterwordstretchfactor\fontdimen3\font minus
  \fontdimen4\font\relax}
\providecommand{\BIBforeignlanguage}[2]{{%
\expandafter\ifx\csname l@#1\endcsname\relax
\typeout{** WARNING: IEEEtran.bst: No hyphenation pattern has been}%
\typeout{** loaded for the language `#1'. Using the pattern for}%
\typeout{** the default language instead.}%
\else
\language=\csname l@#1\endcsname
\fi
#2}}
\providecommand{\BIBdecl}{\relax}
\BIBdecl

\bibitem{rahman2020accountable}
M.~S. Rahman, A.~Al~Omar, M.~Z.~A. Bhuiyan, A.~Basu, S.~Kiyomoto, and G.~Wang,
  ``Accountable cross-border data sharing using blockchain under relaxed trust
  assumption,'' \emph{IEEE Transactions on Engineering Management}, 2020.

\bibitem{al2019privacy}
A.~Al~Omar, M.~Z.~A. Bhuiyan, A.~Basu, S.~Kiyomoto, and M.~S. Rahman,
  ``Privacy-friendly platform for healthcare data in cloud based on blockchain
  environment,'' \emph{Future generation computer systems}, vol.~95, pp.
  511--521, 2019.

\bibitem{nesbitt_ethereum_2019}
\BIBentryALTinterwordspacing
M.~Nesbitt. (2019, Mar.) ``{Ethereum} {Classic} ({ETC}) is currently being 51\%
  attacked''. Accessed: 2020-08-13. [Online]. Available:
  \url{https://tinyurl.com/coinbasedeepEthereum}
\BIBentrySTDinterwordspacing

\bibitem{orcutt_unhackable_2019}
\BIBentryALTinterwordspacing
M.~Orcutt. (2019, Feb.) ``{Once} hailed as unhackable, blockchains are now
  getting hacked''. Accessed: 2020-08-13. [Online]. Available:
  \url{https://tinyurl.com/blockchainHacked}
\BIBentrySTDinterwordspacing

\bibitem{siegel_dao_2016}
\BIBentryALTinterwordspacing
D.~Siegel. (2016, Jun.) ``{Understanding} {The} {DAO} {Attack}''. Accessed:
  2020-08-13. [Online]. Available: \url{https://tinyurl.com/coindeskDAO}
\BIBentrySTDinterwordspacing

\bibitem{adelstein_bitcoin_2016}
\BIBentryALTinterwordspacing
J.~Adelstein and N.-K. Stucky, ``{Inside} the {Biggest} {Bitcoin} {Heist} in
  {History},'' \emph{The Daily Beast}, May 2016, accessed: 2020-08-13.
  [Online]. Available: \url{https://tinyurl.com/dailyBeastMtGox}
\BIBentrySTDinterwordspacing

\bibitem{baldwin_bitfinex_2016}
\BIBentryALTinterwordspacing
C.~Baldwin, ``{Bitcoin} worth \$72 million stolen from {Bitfinex} exchange in
  {Hong} {Kong},'' \emph{Reuters}, Aug. 2016, accessed: 2020-08-13. [Online].
  Available: \url{https://tinyurl.com/reutersBitfinex}
\BIBentrySTDinterwordspacing

\bibitem{noauthor_stride_2019}
\BIBentryALTinterwordspacing
(2019, Aug.) ``{Stride}, {VAST}, {Trike}, \& {More}: {Which} {Threat}
  {Modeling} {Methodology} is {Right} {For} {Your} {Organization}?''. Accessed:
  2020-08-15. [Online]. Available:
  \url{https://tinyurl.com/rightThreatModeling}
\BIBentrySTDinterwordspacing

\bibitem{lipner2006security}
S.~Lipner and M.~Howard, ``The security development lifecycle sdl: A process
  for developing demonstrably more secure software,'' in \emph{IEEE: Annual
  Computer Security Applications Conference}, 2006.

\bibitem{johnstone2010threat}
M.~N. Johnstone, ``Threat modelling with stride and uml,'' 2010.

\bibitem{kohnfelder1999threats}
L.~Kohnfelder and P.~Garg, ``The threats to our products,'' \emph{Microsoft
  Interface, Microsoft Corporation}, vol.~33, 1999.

\bibitem{petri1966communication}
C.~A. Petri, ``Communication with automata,'' 1966.

\bibitem{mcdermott2001attack}
J.~P. McDermott, ``Attack net penetration testing,'' in \emph{Proceedings of
  the 2000 workshop on New security paradigms}, 2001, pp. 15--21.

\bibitem{nakamoto2019bitcoin}
S.~Nakamoto, ``Bitcoin: A peer-to-peer electronic cash system,'' Manubot, Tech.
  Rep., 2019.

\bibitem{szabo1997formalizing}
N.~Szabo, ``Formalizing and securing relationships on public networks,''
  \emph{First Monday}, 1997.

\bibitem{buterin2014next}
V.~Buterin \emph{et~al.}, ``A next-generation smart contract and decentralized
  application platform,'' \emph{white paper}, vol.~3, no.~37, 2014.

\bibitem{ferdous2020blockchain}
M.~S. Ferdous, M.~J.~M. Chowdhury, M.~A. Hoque, and A.~Colman, ``Blockchain
  consensus algorithms: A survey,'' \emph{arXiv preprint arXiv:2001.07091},
  2020.

\bibitem{aaronsonlimits}
S.~Aaronson, ``The limits of quantum computers (draft),'' \emph{Issue of
  scientific American}.

\bibitem{shor1999polynomial}
P.~W. Shor, ``Polynomial-time algorithms for prime factorization and discrete
  logarithms on a quantum computer,'' \emph{SIAM review}, vol.~41, no.~2, pp.
  303--332, 1999.

\bibitem{borbely2007grover}
E.~Borbely, ``Grover search algorithm,'' \emph{arXiv preprint arXiv:0705.4171},
  2007.

\bibitem{saad2019exploring}
M.~Saad, J.~Spaulding, L.~Njilla, C.~Kamhoua, S.~Shetty, D.~Nyang, and
  A.~Mohaisen, ``Exploring the attack surface of blockchain: A systematic
  overview,'' \emph{arXiv preprint arXiv:1904.03487}, 2019.

\bibitem{li2020survey}
X.~Li, P.~Jiang, T.~Chen, X.~Luo, and Q.~Wen, ``A survey on the security of
  blockchain systems,'' \emph{Future Generation Computer Systems}, vol. 107,
  pp. 841--853, 2020.

\bibitem{lee2019systematic}
J.~H. Lee \emph{et~al.}, ``Systematic approach to analyzing security and
  vulnerabilities of blockchain systems,'' Ph.D. dissertation, Massachusetts
  Institute of Technology, 2019.

\bibitem{kabashkin2017risk}
I.~Kabashkin, ``Risk modelling of blockchain ecosystem,'' in
  \emph{International Conference on Network and System Security}.\hskip 1em
  plus 0.5em minus 0.4em\relax Springer, 2017, pp. 59--70.

\bibitem{poston_threat_nodate}
\BIBentryALTinterwordspacing
H.~Poston. ``{Threat} {Modeling} for the {Blockchain}''. Accessed: 2020-08-10.
  [Online]. Available: \url{https://tinyurl.com/hPostonThreat}
\BIBentrySTDinterwordspacing

\bibitem{ferdous2020immutable}
M.~S. Ferdous, M.~J.~M. Chowdhury, K.~Biswas, N.~Chowdhury, and
  V.~Muthukkumarasamy, ``Immutable autobiography of smart cars leveraging
  blockchain technology,'' \emph{Knowledge Engineering Review}, vol.~35, no.~3,
  p.~17, 2020.

\bibitem{anna_blockchain_2018}
\BIBentryALTinterwordspacing
A.~Katrenko and M.~Sotnichek. (2018, Nov.) ``{Blockchain} {Attack} {Vectors}:
  {Vulnerabilities} of the {Most} {Secure} {Technology}''. Accessed:
  2020-08-11. [Online]. Available: \url{https://tinyurl.com/aprioritAttack}
\BIBentrySTDinterwordspacing

\bibitem{chen2011petri}
T.~M. Chen, J.~C. Sanchez-Aarnoutse, and J.~Buford, ``Petri net modeling of
  cyber-physical attacks on smart grid,'' \emph{IEEE Transactions on smart
  grid}, vol.~2, no.~4, pp. 741--749, 2011.

\bibitem{pinna2018petri}
A.~Pinna, R.~Tonelli, M.~Orr{\'u}, and M.~Marchesi, ``A petri nets model for
  blockchain analysis,'' \emph{The Computer Journal}, vol.~61, no.~9, pp.
  1374--1388, 2018.

\bibitem{mayer2016ecdsa}
H.~Mayer, ``{ECDSA} security in bitcoin and ethereum: a research survey,''
  \emph{CoinFaabrik, June}, vol.~28, p. 126, 2016.

\bibitem{kiktenko2018quantum}
E.~O. Kiktenko, N.~O. Pozhar, M.~N. Anufriev, A.~S. Trushechkin, R.~R. Yunusov,
  Y.~V. Kurochkin, A.~Lvovsky, and A.~Fedorov, ``Quantum-secured blockchain,''
  \emph{Quantum Science and Technology}, vol.~3, no.~3, p. 035004, 2018.

\bibitem{natoli2016balance}
C.~Natoli and V.~Gramoli, ``The balance attack against proof-of-work
  blockchains: The r3 testbed as an example,'' \emph{arXiv preprint
  arXiv:1612.09426}, 2016.

\bibitem{atzei2017survey}
N.~Atzei, M.~Bartoletti, and T.~Cimoli, ``A survey of attacks on ethereum smart
  contracts (sok),'' in \emph{International conference on principles of
  security and trust}.\hskip 1em plus 0.5em minus 0.4em\relax Springer, 2017,
  pp. 164--186.

\bibitem{luu2016making}
L.~Luu, D.-H. Chu, H.~Olickel, P.~Saxena, and A.~Hobor, ``Making smart
  contracts smarter,'' in \emph{Proceedings of the 2016 ACM SIGSAC conference
  on computer and communications security}, 2016, pp. 254--269.

\bibitem{buterin_dao_2016}
\BIBentryALTinterwordspacing
V.~Buterin. (2016, Jun.) ``{CRITICAL} {UPDATE} {Re}: {DAO} {Vulnerability}''.
  Accessed: 2020-08-14. [Online]. Available:
  \url{https://tinyurl.com/ethereumDAO}
\BIBentrySTDinterwordspacing

\bibitem{iotex_eclipse_2018}
\BIBentryALTinterwordspacing
(2018, Mar.) ``{Eclipse} {Attacks} on {Blockchains}’ {Peer}-to-{Peer}
  {Network}''. Accessed: 2020-08-11. [Online]. Available:
  \url{https://tinyurl.com/ioTexEclipse}
\BIBentrySTDinterwordspacing

\bibitem{sayeed2019assessing}
S.~Sayeed and H.~Marco-Gisbert, ``Assessing blockchain consensus and security
  mechanisms against the 51\% attack,'' \emph{Applied Sciences}, vol.~9, no.~9,
  p. 1788, 2019.

\bibitem{roetteler2017quantum}
M.~Roetteler, M.~Naehrig, K.~M. Svore, and K.~Lauter, ``Quantum resource
  estimates for computing elliptic curve discrete logarithms,'' in
  \emph{International Conference on the Theory and Application of Cryptology
  and Information Security}.\hskip 1em plus 0.5em minus 0.4em\relax Springer,
  2017, pp. 241--270.

\bibitem{douceur2002sybil}
J.~R. Douceur, ``The sybil attack,'' in \emph{International workshop on
  peer-to-peer systems}.\hskip 1em plus 0.5em minus 0.4em\relax Springer, 2002,
  pp. 251--260.

\bibitem{zhang2019double}
S.~Zhang and J.-H. Lee, ``Double-spending with a sybil attack in the bitcoin
  decentralized network,'' \emph{IEEE Transactions on Industrial Informatics},
  vol.~15, no.~10, pp. 5715--5722, 2019.

\bibitem{swathi2019preventing}
P.~Swathi, C.~Modi, and D.~Patel, ``Preventing sybil attack in blockchain using
  distributed behavior monitoring of miners,'' in \emph{2019 10th International
  Conference on Computing, Communication and Networking Technologies
  (ICCCNT)}.\hskip 1em plus 0.5em minus 0.4em\relax IEEE, 2019, pp. 1--6.

\bibitem{courtois2014subversive}
N.~T. Courtois and L.~Bahack, ``On subversive miner strategies and block
  withholding attack in bitcoin digital currency,'' \emph{arXiv preprint
  arXiv:1402.1718}, 2014.

\bibitem{sapirshtein2016optimal}
A.~Sapirshtein, Y.~Sompolinsky, and A.~Zohar, ``Optimal selfish mining
  strategies in bitcoin,'' in \emph{International Conference on Financial
  Cryptography and Data Security}.\hskip 1em plus 0.5em minus 0.4em\relax
  Springer, 2016, pp. 515--532.

\bibitem{nayak2016stubborn}
K.~Nayak, S.~Kumar, A.~Miller, and E.~Shi, ``Stubborn mining: Generalizing
  selfish mining and combining with an eclipse attack,'' in \emph{2016 IEEE
  European Symposium on Security and Privacy (EuroS\&P)}.\hskip 1em plus 0.5em
  minus 0.4em\relax IEEE, 2016, pp. 305--320.

\bibitem{karame2012double}
G.~O. Karame, E.~Androulaki, and S.~Capkun, ``Double-spending fast payments in
  bitcoin,'' in \emph{Proceedings of the 2012 ACM conference on Computer and
  communications security}, 2012, pp. 906--917.

\bibitem{rosenfeld2014analysis}
M.~Rosenfeld, ``Analysis of hashrate-based double spending,'' \emph{arXiv
  preprint arXiv:1402.2009}, 2014.

\bibitem{hal_finney_2011}
\BIBentryALTinterwordspacing
Hal. (2011, Feb.) ``{Best} practice for fast transaction acceptance - how high
  is the risk?''. Accessed: 2020-08-11. [Online]. Available:
  \url{https://tinyurl.com/bitcoinFinney}
\BIBentrySTDinterwordspacing

\bibitem{bitstack_finney_nodate}
\BIBentryALTinterwordspacing
``{What} is a {Finney} attack?''. Accessed: 2020-08-11. [Online]. Available:
  \url{https://tinyurl.com/stackFinney}
\BIBentrySTDinterwordspacing

\bibitem{baqer2016stressing}
K.~Baqer, D.~Y. Huang, D.~McCoy, and N.~Weaver, ``Stressing out: Bitcoin
  “stress testing”,'' in \emph{International Conference on Financial
  Cryptography and Data Security}.\hskip 1em plus 0.5em minus 0.4em\relax
  Springer, 2016, pp. 3--18.

\bibitem{gao2001inferring}
L.~Gao, ``On inferring autonomous system relationships in the internet,''
  \emph{IEEE/ACM Transactions on networking}, vol.~9, no.~6, pp. 733--745,
  2001.

\bibitem{rosenfeld2011analysis}
M.~Rosenfeld, ``Analysis of bitcoin pooled mining reward systems,'' \emph{arXiv
  preprint arXiv:1112.4980}, 2011.

\bibitem{abdullah2017blockchain}
N.~Abdullah, A.~Hakansson, and E.~Moradian, ``Blockchain based approach to
  enhance big data authentication in distributed environment,'' in \emph{ICUFN'
  2017}.\hskip 1em plus 0.5em minus 0.4em\relax IEEE, 2017, pp. 887--892.

\end{thebibliography}

\end{document}